\documentclass[a4paper,11pt]{article}
\usepackage{pos}
\usepackage{amsmath,graphicx,multirow,tabularx,physics}
\usepackage{slashed,listings}
\usepackage{float}
\usepackage{wrapfig}

\definecolor{airforceblue}{rgb}{0.36,0.54,0.66}
\definecolor{burgundy}{rgb}{0.5,0.0,0.13}
\definecolor{blue-violet}{rgb}{0.54,0.17,0.89}
\title{Fourier coefficients of the net-baryon number density}

\ShortTitle{Fourier coefficients of the net-baryon number density}
\author*[a]{Christian Schmidt}
\affiliation[a]{Fakultät für Physik, Universität Bielefeld,\\
Universitätsstraße 25, Bielefeld, Germany} 
\emailAdd{schmidt@physik.uni-bielefeld.de}
\abstract{We calculate Fourier coefficients of the net-baryon number as a function of a purely imaginary chemical potential. The asymptotic behavior of
these coefficients is governed by the singularity structure of the QCD
partition function and thus encodes information on phase
transitions. For the calculation of the Fourier
coefficients from lattice data of the Bielefeld-Parma collaboration we use a novel Filon-type quadrature, designed for highly oscillatory integrals. We find sensitivity to chiral scaling in a narrow temperature interval below the Roberge-Weiss transition temperature. Scaling fits yield reasonable values for the position of the Lee-Yang edge singularity in the complex chemical potential
plane. Our lattice data has been obtained from simulations with
(2+1)-flavors of highly improved staggered quarks (HISQ) at imaginary
chemical potential on $N_\tau=4, 6$ and $8$ lattices at physical quark
masses.}

\FullConference{The 39th International Symposium on Lattice Field Theory,\\
8th-13th August, 2022,\\
Rheinische Friedrich-Wilhelms-Universität Bonn, Bonn, Germany}

\begin{document}
\maketitle
\section{Introduction}
A detailed calculation of the QCD phase diagram at nonzero temperature and density from first principles is an unsolved open issue. 
Unfortunately, lattice QCD calculations are hindered by the infamous sign problem as soon as a non-vanishing chemical potential $\mu>0$ is introduced.
In order to alleviate or circumvent the sign problem, many numerical methods have been developed, which include the Taylor expansion about $\mu=0$ as well as calculations at purely imaginary chemical potential $\mu=i\mu^I$, combined with an analytic continuation to real $\mu$ values. 
Since the QCD partition function is periodic in $\mu^I/T\equiv\hat\mu^I$ \cite{Roberge:1986mm}, with periodicity $2\pi/N_c$, where $N_c$ denotes the number of colors, it is quite natural to analyze data that is obtained from lattice QCD calculations with imaginary chemical potential in terms of a Fourier expansion. 
In particular, the Fourier decomposition of the net baryon number density,  
\begin{equation}
    \chi_1^B(T,\hat\mu_B)=\frac{n_B(T,\hat\mu_B)}{T^3}=\frac{1}{VT^3}\frac{\partial}{\hat\mu_B} \ln Z_{GC}(T,\hat\mu_B),
\end{equation}
where $Z_{GC}(T,\hat\mu_B )$ is the grand canonical partition function\footnote{The dependence on the volume V is suppressed for simplicity.}, $\hat\mu_B = \mu_B /T$ is the reduced baryon chemical potential, and $T$ the temperature, is straightforwardly accessible from lattice QCD data and has been the starting point of many recent studies \cite{Vovchenko:2017xad, Vovchenko:2017gkg, Almasi:2018lok, Almasi:2019bvl, Bornyakov:2016wld, Bornyakov:2022blw}. 
As $\chi_1^B$ exhibits the same periodicity as the partition function and in addition is an odd function of $\mu_B^I$, we can expand $\text{Im}\chi_1^B$ as a Fourier sine series 
\begin{equation}
    \text{Re}[\chi_1^B(T,i\hat\mu_B^I)]=0, \quad \text{Im}[\chi_1^B(T,i\hat\mu_B^I)]=\sum_{k=1}^{\infty}b_k(T)\sin(k\hat\mu_B^I).
    \label{eq:sine-expanison}
\end{equation}
Note that the real part of $\chi_1^B$ vanishes identically at $\hat\mu_B=i\hat\mu_B^I$. The set of coefficients $\{b_k(T)\}_{k=1}^{\infty}$ encode -- up to an unimportant integration constant -- the complete information on the QCD partition function and thus on the thermodynamic properties of QCD matter in the region of the phase diagram where the series,  Eq.~(\ref{eq:sine-expanison}), converges. 
This is similar to the set of Taylor expansion coefficients of the dimensionless pressure $\{c_{2k}(T)\}_{k=0}^{\infty}$, defined as
\begin{equation}
    \frac{p(T,\hat\mu_B)}{T^4}=\frac{1}{VT^3}\ln Z_{GC}(T,\hat\mu_B)=\sum_{k=0}^{\infty}c_{2k}(T)\hat\mu_B^{2k}.
\end{equation}
The calculation of the coefficients $c_{2k}(T)$ is numerically very demanding and currently only $c_{2k}(T)$ for $k\le 4$ are known from direct calculations at $\mu_B=0$. 
Moreover, statistical and systematical errors for $c_6(T)$ and $c_8(T)$ are still very large, for recent results see \cite{EoS2022}. 
It might thus be tempting to verify or even complement the information from the known Taylor coefficients by a calculation of the first few Fourier coefficients $b_k(T)$. 
Unfortunately, the calculation of the coefficients $b_k(T)$ is equally difficult. 
However, just as the Taylor coefficients have a physical interpretation as cumulants of the net baryon charge $B$, the coefficients $b_k(T)$ bear some physical meaning as well. The Fourier expansion can formally be seen as a fugacity expansion. 
The set of available $\{b_k(T)\}$ can thus be used to determine the canonical partition functions $Z_C(T,V,N)$ \cite{Bornyakov:2016wld, Bornyakov:2022blw}.
For the same reason the Fourier expansion of the net baryon number density can be understood as a relativistic extension of Mayer’s cluster expansion in fugacities \cite{Vovchenko:2017gkg}.
In that spirit, the first coefficient $b_1(T)$ is given by the partial pressure of the (non interacting) $|B|=1$ sector, whereas $b_2(T)$ parametrizes the leading order of the baryon-baryon interaction. 
Based on the cluster expansion, the authors of Ref.~\cite{Vovchenko:2017gkg} have introduced a model (CEM) that can predict the coefficients $\{b_k(T)\}_{k=2}^{\infty}$, based on the first two coefficients $b_1(T)$ and $b_2(T)$. 
While this model verifies lattice results for the coefficients $b_3(T)$ and $b_4(T)$, it exhibits an exponential decay of $b_k(T)$ for $k\to\infty$ at fixed $T$ and thus does not incorporate critical behavior. 
The asymptotic behavior of the model was adjusted to a power-law decay in Ref.~\cite{Almasi:2018lok} (RFM), without spoiling the agreement with the lattice data. 
A more thorough investigation of the asymptotic behavior of the $\{b_k(T)\}$ in terms of the universal $O(4)$-critical scaling was performed in Ref.~\cite{Almasi:2019bvl}.

We are aiming at an analysis of the analytic structure of the QCD phase diagram, by means of Lee-Yang zeros \cite{Yang:1952be}. 
Zeros of the partition function will manifest as poles of the baryon number density $\chi_1^B(T)$ in the complex $\hat\mu_B$ plane. 
A new method for the analytic continuation of $\chi_1^B(T)$ was introduced in Ref.~\cite{Dimopoulos:2021vrk} and is based on a multi-point Padé analysis.
Since this method yields a rational function approximation to the lattice data, it is straightforward to determine poles of the observable in the complex $\mu_B$ plane. 
The closest pole might be associated with the Lee-Yang edge singularity in QCD, which exhibits a well defined universal scaling and can be used to determine various non-universal parameters, including the location of the critical point. 
Recent results on Lee-Yang edge singularities and their scaling have been presented on this conference \cite{Francesco_latt22, Kevin_latt22}.
The verification of this method has been demonstrated by considering the universal scaling in the vicinity of the Roberge-Weiss transition in QCD \cite{Dimopoulos:2021vrk, Kevin_latt22, Schmidt:2022ogw}. 
Here we will introduce a new method for the numerical calculation of the Fourier coefficients $\{b_k(T)\}$ and  verify expected signals on universal scaling in their large $k$ behavior.

\section{Determination of the Fourier coefficients}
In many applications the Fourier coefficients are calculated by the conventional Discrete-Fourier-Transform (DFT) or the popular Fast-Fourier-Transformation (FFT) algorithms. 
However, as our input data stems from lattice QCD calculations and we are interested in the large $k$ behavior of $\{b_k(T)\}$, these algorithms have two crucial drawbacks for our purpose. 
Firstly, the number of detectable frequencies is directly related to the sampling rate of the function. However, lattice QCD calculations are expensive and we want to keep the number of sampling points $N$ low. 
Secondly, the numerical (root-mean-square) error of the DFT and FFT algorithm increases at least as $\sim N^{1/2}$ \cite{doi:10.1137/S1064827593247023}. 
It thus seems advantageous to first perform a numerical interpolation of the lattice data before calculating the Fourier coefficients. 
Furthermore, it is obvious that the calculation of $b_k(T)$ demands solving a highly oscillatory integral, we have 
\begin{equation}
    b_k(T)=\frac{2}{\pi}\int\limits_0^{\pi}\text{Im}\left[\chi_1^B\left(T,i\hat\mu_B^I\right)\right]\sin(k\hat\mu_B^I)\;\text{d}\hat\mu_B^I.
    \label{eq:bk}
\end{equation}
A popular numerical method for oscillatory integrals is the Filon-type quadrature, which simply makes use of the interpolating polynomial for the non oscillatory part of the integrand (here $\text{Im}[\chi_1^B(T,i\hat\mu_B^I)]$), whereas the oscillator (here $\sin(k\hat\mu_B^I)$) is treated analytically. 
This method has the advantage that it is asymptotic in the sense that its error decreases with increasing frequency $k$. 
In Ref.~\cite{Iserles:2004} a Filon-type quadrature has been constructed which uses in addition to the interpolating data also derivatives, i.e. the interpolating polynomial is taken to be a piece-wise polynomial, matching values and derivatives to order $s$ (Hermite interpolation) at the boundaries.  
The asymptotic analysis performed in \cite{Iserles:2004} shows that for this method, and a general oscillator of the form $e^{i\omega g(x)}$, the error decreases as $\mathcal{O}(\omega^{-s-2})$.

\begin{figure}
    \centering
    \includegraphics[width=\textwidth]{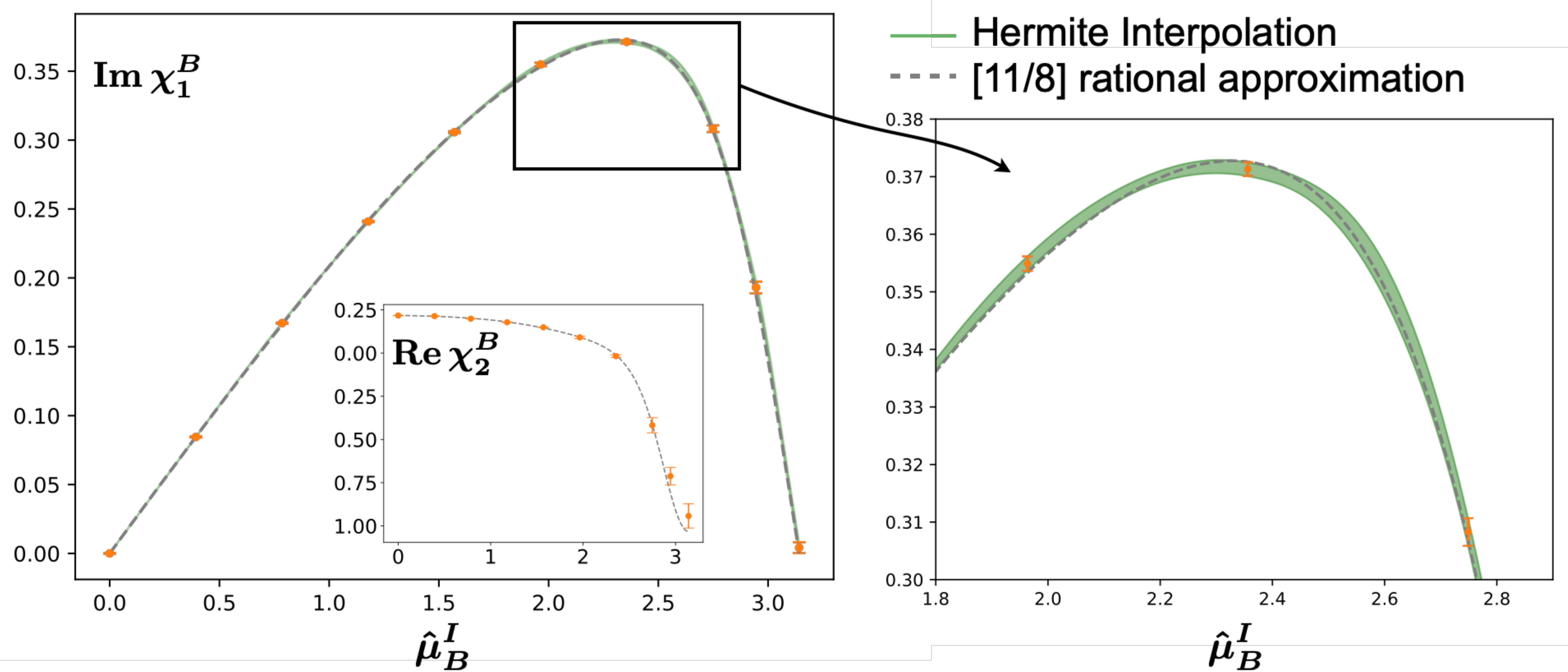}
    \caption{Comparison of the Hermite interpolation of the net baryon number density $\text{Im}[\chi_1^B(T,i\hat\mu_B^I$)] as function of $\mu_B^I$ with a [11/8] rational approximation. The lattice data is obtained from a calculation on a $36^3\times6$ lattice at $T=190$ MeV using SIMULATeQCD \cite{Bollweg:2021cvl}. The inlay on the left shows the second baryon number cumulant $\text{Re}[\chi_2^B(T,i\hat\mu_B^I)]$, together with the first derivative of the rational approximation of $\text{Im}[\chi_1^B(T,i\hat\mu_B^I)]$. The figure on the right is a zoom into the peak region.  }
    \label{fig:Hermite}
\end{figure}
\begin{figure}
    \centering
    \includegraphics[width=\textwidth]{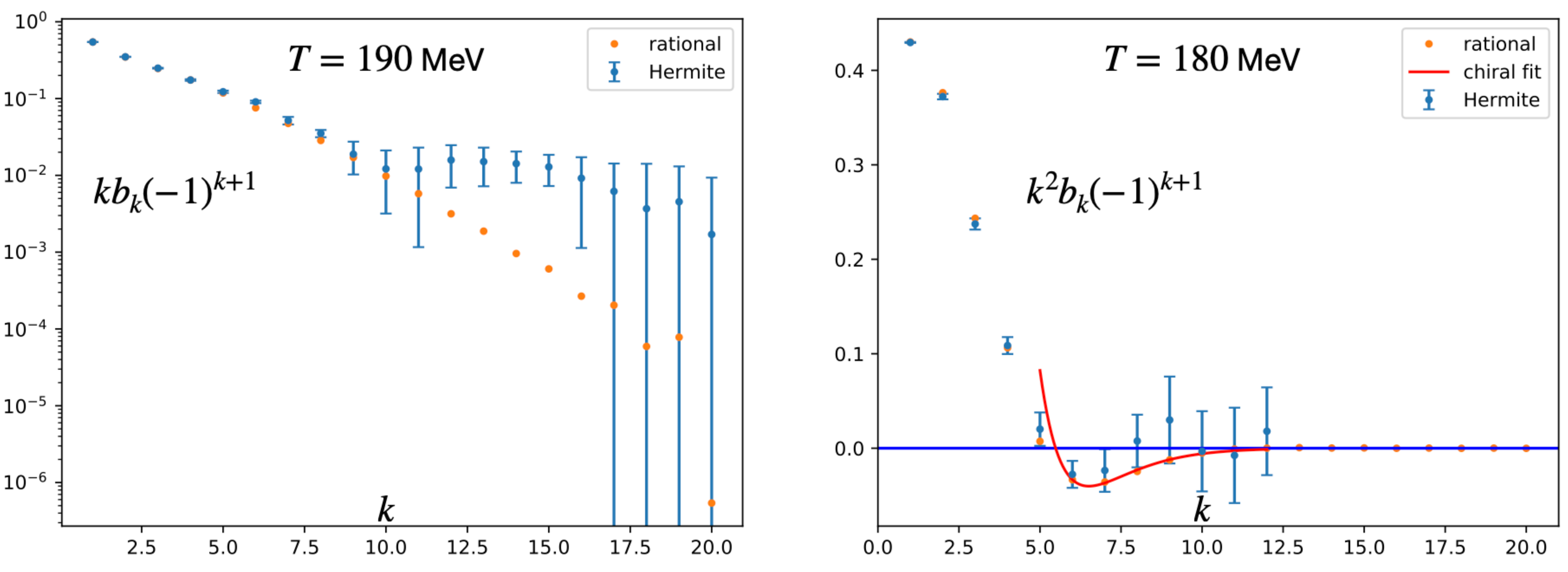}
    \caption{Preliminary calculation of Fourier coefficients $b_k(T)$ of the net baryon number density $\text{Im}[\chi_1^B]$ as a function of the frequency $k$ obtained from an analytic integration based on two different interpolating functions (Hermite and rational). The left (right) panel shows results for $T=190$ ($T=180$) MeV. The normalizing factor is $k(-1)^{k+1}$ (left) and $k^2(-1)^{k+1}$ (right). The calculation is based on lattice data of $\chi_1^B$, $\chi_2^B$ and $\chi_3^B$ from $36^3\times 6$ lattices using SIMULATeQCD \cite{Bollweg:2021cvl}. A Fit to ansatz Eq.~(\ref{eq:chiral}) to the data at $T=180$ MeV is also shown.}
    \label{fig:bk}
\end{figure}

In Fig.~\ref{fig:Hermite} we show the $2^{nd}$ order Hermite interpolation to $\text{Im}[\chi_1^B(T,i\hat\mu_B^I$)], which continuously matches $\chi_1^B$, $\chi_2^B$ and $\chi_3^B$. 
The green error band is obtained by assuming independent, normally distributed errors at the simulation points and bootstrapping. 
For comparison we show a [11/8] rational polynomial obtained with the multi-point Padé method \cite{Dimopoulos:2021vrk}. 
We see that both methods always agree within errors, even in the peak region where the differences are most pronounced. 
Once we have an analytic expression for an interpolating function at hand, it is easy to calculate the integral Eq.~(\ref{eq:bk}) analytically. 
In particular, for the Hermite interpolation we can split the integration over the interval $[0,\pi]$ to a sum over the intervals defined by the $N$ data points. We have
\begin{equation}
    b_k(T)=\sum_{j=1}^{N-1}\int\limits_{\hat\mu_B^{(j)}}^{\hat\mu_B^{(j+1)}}p_j(x)\sin(kx)\;\text{d}x,\quad\text{with}\quad 0=\hat\mu_B^{1}<\hat\mu_B^{2}\cdots<\hat\mu_B^{N}=\pi
\end{equation}
denoting the $N$ locations of the data points and $p_j(x)$ the interpolating polynomial of $\text{Im}[\chi_1^B(T,ix)]$ for $x\in[\hat\mu_B^{(j)}, \hat\mu_B^{(j+1)}]$. 
The results of the analytic integration for both types of interpolations are shown in Fig.~\ref{fig:bk}.
The left (right) panel shows the results for $T=190$ ($T=180$) MeV. 
Error bars for the Hermite interpolation are again obtained from bootstrapping. 
We find that both interpolations yield consistent results at least up to frequency $k\lesssim 10$. 

\section{Universal scaling}
Finally we discuss the expected asymptotic behavior of the Fourier coefficients $b_k(T)$ in the vicinity of a phase transition \cite{Almasi:2019bvl}. 
For $O(N)$ and $Z(2)$ symmetric spin models in 3d, it is well known that the order parameter $M\sim\partial \ln Z(T,h)/\partial h$, where $h$ is the symmetry breaking field, exhibits branch-cuts in the complex $h$-plane. 
The position of the branch-cut singularity is identical to the Lee-yang edge (LYE) singularity, defined as the point where the linear density of the Lee-Yang zeros diverges in the continuum limit. 
For the analysis here, we estimate the leading singular behavior of the net baryon number density $\chi_1^B$. This is particularly easy in case of the Roberge-Weiss transition, where we find $\text{Im}[\chi_1^B] \sim M$ and $h \sim \mu_B^I$. For fixed $T=T_{RW}$ we thus assume $\text{Im}[\chi_1^B] \sim (\pi-\hat\mu_B^I)^{1/\delta}$, where $\delta$ refers to a a critical exponent of the 3d Z(2) universality class. 
For the Fourier coefficients one thus obtains
\begin{equation}
    b_k\sim\int\limits_0^\pi\text{d}\hat\mu_B^I\;(\pi-\hat\mu_B^I)^{1/\delta}\sin(k\hat\mu_B^I)\sim\frac{(-1)^{k+1}}{k^{1+1/\delta}}.
\end{equation}
The analysis is similar but more involved in the case of the chiral O(4) transition in presence of an explicit symmetry breaking quark mass (crossover). 
In essence one finds \cite{Almasi:2019bvl}
\begin{equation}
    b_k\sim \frac{e^{-k\hat\mu_{LYE}^R}}{k^{2-\alpha}}\left(\sin(k\hat\mu_{LYE}^I-\alpha\pi/2)+R_\pm\sin(k\hat\mu_{LYE}^I+\alpha\pi/2)\right),
    \label{eq:chiral}
\end{equation}
for $T_{cep}<T<T_{RW}$. Here $T_{cep}$ denotes the temperature of the QCD critical point, the branch-cut singularity is located at $\hat\mu_{LYE}=\hat\mu_{LYE}^R+i\hat\mu_{LYE}^I$, and $\alpha\approx-0.21$ and $R_\pm\approx 1.85$ denote universal quantities from the O(4) universality class. 
Hence, the behavior resembles a damped oscillation were the exponential suppression relates to the real part of the LYE and the period of the oscillation to the imaginary part of the LYE. 

In Fig.~\ref{fig:bk} we show examples for both of these scenarios. However, a clear oscillatory behavior, which indicates sensitivity to the chiral O(4) transition could only be found for two of our temperatures, $T=180$ and $T=185$ MeV. 
For $T<185$ MeV, the suppression due to the real part is so large that the oscillations are hidden in the noise. 
Fits to the asymptotic behavior of $b_k(T)$, for $T=180$ and $T=185$ MeV with ansatz Eq.~(\ref{eq:chiral}) yield locations for the LYE which are consistent with results from the poles of the multi-point Padé \cite{Kevin_latt22}. 
In fact, the real parts are in good agreement, the imaginary parts come out slightly lower. The fit shown in Fig.~\ref{fig:bk} yields $\hat\mu_{LYE}=0.97(6)+3.123(3)i$.

\section{Summary, conclusion and outlook}
We have presented a preliminary calculation of Fourier coefficients $\{b_k(T)\}$ of the net baryon number $\text{Im}[\chi_1^B(T,i\hat\mu_B^I)]$.
The calculation is based on lattice data from the Bielefeld-Parma collaboration \cite{Kevin_latt22} and uses a novel Filon-type quadrature. 
With this method we were able to obtain Fourier coefficients for frequencies of $k\lesssim 10$. 
Through the asymptotic behavior of these coefficients one might identify branch-cut singularities in the complex chemical potential plane. 
However, sensitivity to the chiral O(4) transition was only found in a narrow temperature interval $T\in[180,185]$ MeV. 
For temperatures below $T=180$ MeV, the exponential suppression with is associated with the real part of the LYE seems too strong. 
To alleviate this problem in future calculation we might improve the numerical quadrature further by investigate adaptive Filon-type methods and go to lighter than physical quark masses. 
The latter will reduce the real part of the LYE and thus lift the exponential suppression.

\section*{Acknowledgments}
This work was supported in part by the Deutsche Forschungsgemeinschaft (DFG) through the grant 315477589-TRR 211 and "NFDI 39/1" for the PUNCH4NFDI consortium and the grant EU H2020-MSCA-ITN-2018-813942 (EuroPLEx) of the European Union.

\bibliographystyle{JHEP} 
\bibliography{bib}

\end{document}